\documentclass[11pt,a4paper]{article}
\usepackage[margin=2.5cm]{geometry}
\usepackage[nocompress]{cite}

\usepackage{booktabs}

\usepackage{amsmath}
\usepackage{amssymb}

\usepackage{graphicx}
\usepackage{subcaption}
\usepackage{pgfplots}
\usepgfplotslibrary{groupplots}
\pgfplotsset{compat=1.18}
\usepackage{xcolor}
\usepackage{multirow}
\usepackage{rotating}
\usepackage{float}

\usepackage{hyperref}

\begin{document}
\definecolor{brightpink}{rgb}{1.0, 0.0, 0.5}
\newcommand{\ngc}[1]{{\color{brightpink} (\textbf{NG:} #1)}}
\newcommand{\ngi}[1]{{{\color{brightpink} #1}}}

\definecolor{forestmydarkgreen(web)}{rgb}{0.13, 0.55, 0.13}
\newcommand{\av}[1]{{\color{forestmydarkgreen(web)} (\textbf{AV:} #1)}}

\definecolor{skyblue}{RGB}{51,153,255}
\newcommand{\ale}[1]{{\color{skyblue} (\textbf{Alex:} #1)}}
\definecolor{darkgreen}{rgb}{0.13, 0.55, 0.13}

\title{Matrix Factorization Framework for Community Detection\\ under the Degree-Corrected Block Model}

\author{%
Alexandra Dache, Arnaud Vandaele and Nicolas Gillis\\[0.5em]
{\small Department of Mathematics and Operational Research}\\
{\small University of Mons, Mons, Belgium}\\
{\small\texttt{firstname.lastname@umons.ac.be}}
}

\date{} 

\maketitle

\begin{abstract} 
 Community detection is a fundamental task in data analysis, and block models provide an approach for identifying diverse community structures. The degree-corrected block model (DCBM) is an established model that captures node degree heterogeneity. However, inference methods are computationally costly and sensitive to initialization, while cheaper alternatives, such as spectral or modularity-based approaches, may be less effective at recovering certain non-assortative structures. In this work, we show that DCBM inference can be reformulated as a constrained nonnegative matrix trifactorization problem. Leveraging this insight, we propose a scalable community detection method with a theoretically grounded initialization providing an initial partition for inference algorithms. Our approach is agnostic to network structure and applies to any graph representable by a DCBM. 
 Experiments on synthetic and real-world benchmark networks show that our method achieves community recovery comparable to, and sometimes better than, that achieved by the DCBM inference methods, while being faster: it processes a graph with 100,000 nodes and 500,000 edges in about 30 seconds.
 Moreover, the initialization significantly improves solution quality. 
 Overall, this work provides a scalable and robust framework for community detection and highlights the benefits of a matrix-factorization perspective for the DCBM. 
\end{abstract}


\section{Introduction}
\begin{figure}[]
\centering
\hspace*{\fill}
\begin{subfigure}{0.2\textwidth}
    \centering
    \includegraphics[width=1\linewidth]{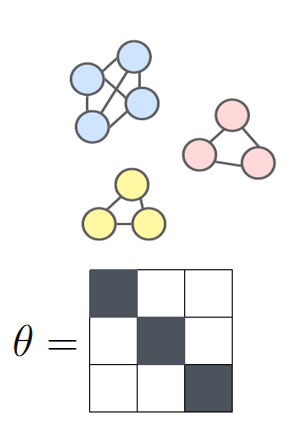}
    \caption{Assortative}
    \label{fig:assortative}   
\end{subfigure}
\hspace{0.3cm}
\begin{subfigure}{0.2\textwidth}
    \centering
     \includegraphics[width=1\linewidth]{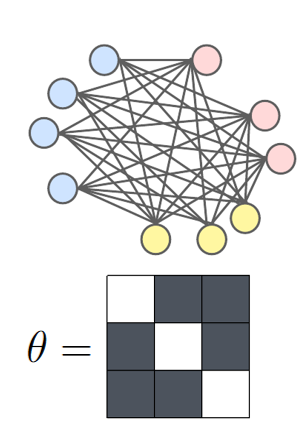}
    \caption{Disassortative}
    \label{fig:disassortative} 
\end{subfigure}
\hspace{0.3cm}
\begin{subfigure}{0.2\textwidth}
    \centering
    \includegraphics[width=1\linewidth]{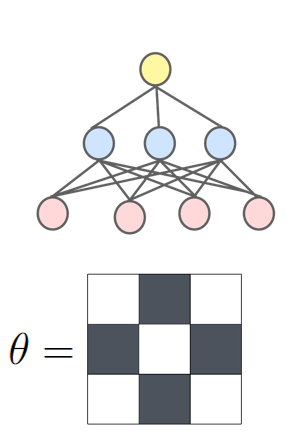}
    \caption{Chain-like}
    \label{fig:chainlike}
\end{subfigure}
\hspace*{\fill}
\caption{
Examples of structures that can be represented by an SBM with three blocks, illustrated with the matrix $\theta$, where entries with high values are shown in black. 
\label{fig:structures}
}
\end{figure}

 The stochastic block model (SBM), introduced by~\cite{SBM_first}, models a network with blocks of nodes, where the probability of an interaction between two nodes depends only on the blocks to which they belong. An SBM with $n$ nodes divided into $r$ blocks or communities can be fully characterized by two parameter matrices. The first is an $n \times r$ matrix, denoted $Z$, which encodes the community to which each node belongs: $Z(i, k) = 1$ if node $i$ is assigned to community $k$, and $Z(i, k) = 0$ otherwise. The second is an $r \times r$ matrix of probabilities, denoted $\theta$, where $\theta(k, l)$ represents the probability that an edge exists between a node belonging to community $k$ and a node belonging to community $l$. An undirected graph with an adjacency matrix $A$ follows an SBM if $A(i,j) = A(j,i) \sim \text{Bernoulli}((Z\theta Z^\top)_{i,j})$, where each edge is distributed according to a Bernoulli distribution given the communities of the nodes. Given $Z$ and $\theta$, the likelihood of observing the adjacency matrix $A$ is
\begin{equation}
P(A|Z,\theta) = \prod_{j < i}^n (Z\theta Z^\top)_{i,j}^{A_{i,j}} (1 - (Z\theta Z^\top)_{i,j})^{(1 - A_{i,j})}.
\label{eq: Bernoulli DCBM}
\end{equation}
A crucial task, called inference, consists of estimating the most probable parameters $Z$ and $\theta$ from the adjacency matrix in order to maximize this likelihood. SBMs owe their success to their simplicity and the variety of network structures they can model. Indeed, unlike most methods that only identify assortative structures, where nodes are more densely connected within the same community than between communities~\cite{CommunityDetection_survey}, SBMs can identify a wide range of structures and combinations of these; see Fig.~\ref{fig:structures} for an illustration. 

SBMs have been widely studied~\cite{SBM_superreview,SBMasymptotic}. However, this model has a major limitation: it assumes that within a community, all nodes have the same importance and the same connection probabilities, which leads to an identical degree distribution for all nodes within the same community. In real networks, however, node degrees are typically heterogeneous. As a result, the classic SBM struggles to detect communities and tends to group nodes primarily based on their number of connections~\cite{SBM_degreecorection}. The paper~\cite{model_selection} proposes a rigorous method for determining if the degree heterogeneity is too high to use a standard SBM. To account for degree heterogeneity, the degree-corrected block model (DCBM)~\cite{SBM_degreecorection} allows the matrix $Z$ to have non-binary weights in $[0,1]$: each row of $Z$ has at most one nonzero entry, located in the column corresponding to the node's community, and this entry takes a value in $[0,1]$.  The probability of having an edge between nodes $i$ and $j$, which belong to communities $k_i$ and $k_j$, is then given by $Z(i,k_i)\, 
\theta(k_i,k_j)\, Z(j,k_j)$, where $Z(i,k_i)$ can be interpreted as the sociability level of node $i$. The larger $Z(i,k_i)$, the more likely node $i$ is to be connected to other nodes. To simplify calculations, the DCBM of Karrer and Newman~\cite{SBM_degreecorection} allows self-edges and approximates the Bernoulli distribution with a Poisson distribution:
\begin{equation} \label{eq:DCBM-Poisson}
A_{i,j} = A_{j,i} \; \sim \; \text{Poisson}\left( \left( Z \theta Z^\top \right)_{i,j} \right), 
\end{equation}
which incorporates the possibility of multi-edges. The parameter $\theta(k,l)$, originally representing the probability of an edge between communities $k$ and $l$, is now the expected number of edges between them. Note that by convention, for a multi-graph, $A_{i,j}$ is equal to the number of edges between nodes $i$ and $j$ when $i\neq j$, but the diagonal element $A_{i,i}$ is equal to twice the number of self-edges from $i$ to itself. The likelihood of observing the adjacency matrix $A$ given $Z$ and $\theta$ is then:
\begin{align}
P(A|Z,\theta)  = \prod_{j < i}^n \frac{\left(\left(Z\theta Z^\top\right)_{i,j}\right)^{A_{i,j}}}{A_{i,j}!}\exp \left(-\left(Z\theta Z^\top\right)_{i,j}\right)\notag \\
 \quad \quad \times\prod_{i}^n \frac{\left(\frac{1}{2}\left(Z\theta Z^\top\right)_{i,i}\right)^{A_{i,i}/2}}{(A_{i,i}/2)!}\exp \left(-\frac{1}{2}\left(Z\theta Z^\top\right)_{i,i}\right).
\label{eq: DCBM proba}
\end{align}
For sparse networks, the Poisson distribution differs negligibly from the Bernoulli distribution, as the probability of an edge and the expected number of edges become very close. An advantage of modeling with Poisson is that, for a fixed partition of nodes into communities, the parameters $Z$ and $\theta$ that maximize the likelihood~\eqref{eq: DCBM proba} can be computed in closed form~\cite{SBM_degreecorection}. 
Hence, the task reduces to finding the partition that maximizes the likelihood~\eqref{eq: DCBM proba}. By substituting the closed-form expressions of $Z$ and $\theta$ and taking the logarithm, Karrer and Newman~\cite{SBM_degreecorection} showed that maximizing~\eqref{eq: DCBM proba} is equivalent to finding a partition $\mathcal{P}$ that maximizes the unnormalized log-likelihood: 
\begin{equation}   \label{eq: obj_DCBM} 
  \mathcal{L}(A|\mathcal{P})=\sum_{k=1}^r \sum_{l=1}^r  m_{kl}\log\frac{m_{kl}}{\kappa_k\kappa_l}, 
\end{equation} 
where $m_{kl}=\sum_{i,j}A_{ij} \, \delta_{k_i,k} \, \delta_{k_j,l}$ is the total number of edges between communities $k$ and $l$ (counted twice if \mbox{$k=l$}), $\delta_{k_i,k}$ is  the Kronecker delta function, $k_{i}$ denotes the community assignment of node $i$, and $\kappa_k=\sum_l m_{kl}$ is the sum of the degrees of nodes in community $k$. 

To estimate a good partition of size $r$ from the graph, 
there are numerous heuristics that, starting from an initial partition, perform node moves to maximize the log-likelihood~\eqref{eq: obj_DCBM}. The computational cost of computing the change in the log-likelihood when moving a node to another community is $\mathcal{O}(\min(r,\langle d \rangle)+\langle d \rangle)$, where $\langle d \rangle$ is the average node degree of the graph. For large sparse graphs, where $\langle d \rangle \ll r$, the cost reduces to $\mathcal{O}(\langle d \rangle)$, depending only on the average degree. The initial partition is generated randomly and may sometimes be bad, which can lead the algorithm to converge to poor local maxima, as pointed out by~\cite{SCORE,peixoto2014efficient}. To address this, the algorithm can be executed multiple times with different random initializations, with the final partition chosen as the one that maximizes the log-likelihood. Peixoto~\cite{peixoto2014efficient} proposed an alternative based on a greedy agglomerative procedure: starting with each node in its own community, the algorithm progressively merges communities and refines the partition through individual node moves. Spectral and modularity-based methods offer alternative approaches to community detection~\cite{SCORE,Polynom_assos,modmaxforDCBM}. Here, we focus on inference under general block models, which accommodate diverse graph structures without imposing assortativity (e.g., assortative, disassortative, mixed, bipartite, unipartite; see Fig.~\ref{fig:structures}). Our method preserves this flexibility. We therefore focus our comparison on likelihood-based inference methods that can recover different block structures without specifying the type of structure in advance. Throughout, we assume that the number of communities, $r$, is known. An approach to estimate $r$ is to evaluate the model for multiple values of $r$ and retain the one that optimizes a criterion reflecting a trade-off between the number of model parameters and the goodness of fit. We leave model selection outside the scope of this work. 

In this work, we establish a formal link between the DCBM and matrix factorization. We show that inferring the DCBM proposed by Karrer and Newman~\cite{SBM_degreecorection} is equivalent to solving a constrained nonnegative matrix trifactorization problem under the KL divergence. Motivated by the limitations of the KL divergence, we investigate OtrisymNMF as an alternative, in which the KL divergence is replaced by the Frobenius norm, and we present FROST, an efficient algorithm to infer this model. Building on the matrix-factorization perspective, we propose a theoretically-grounded method for obtaining initial estimates of the parameters $Z$ and $\theta$, independently of the graph structure. These estimates serve as starting points for FROST and for inference methods for the DCBM. Finally, experiments on synthetic and real-world networks show that OtrisymNMF, with its inference method FROST, achieves community-detection performance comparable to that of the DCBM methods and, in practice, is faster and, in some cases, more accurate.

\section{A Novel Approach Based on Matrix Factorization}
 
Inference of a DCBM can be formulated as a matrix factorization problem. Given the adjacency matrix of an undirected graph, $A \in \{0,1\}^{n \times n}$, and a number of communities, $r$, we seek to solve:
\begin{equation}
\min_{Z \in \mathbb{R}^{n \times r}_+, 
\theta \in \mathbb{R}^{r \times r}_+} 
d( A , Z \theta Z^\top ) 
\; \text{ s.t. } \; Z^\top Z = I_r,\; \theta^\top=\theta,
\label{eq: factomat}
\end{equation}
where $d(A,B)$ measures the discrepancy between matrices $A$ and $B$, and $I_r$ is the identity matrix of dimension $r$. Both matrices $Z$ and $\theta$ are componentwise nonnegative, and $\theta$ is additionally constrained to be symmetric. The orthogonality constraint $Z^\top Z=I_r$, together with the nonnegativity of $Z$, guarantees that the columns of $Z$ have disjoint supports, ensuring non-overlapping communities. It also normalizes each column to unit $\ell_2$ norm. This normalization can be performed without loss of generality since the columns of $Z$ are determined only up to a multiplicative constant, which can be absorbed into the corresponding rows and columns of $\theta$. Indeed, we can multiply $Z$ by a diagonal matrix $C$ while preserving both the support of $Z$ and the product $Z\theta Z^\top$ (up to an appropriate transformation of $\theta$):
\begin{equation*}
    Z\theta Z^\top=(ZC)(C^{-1}\theta C^{-1})(ZC)^{\top}.
\end{equation*}
Note that in the original DCBM formulation~\cite{SBM_degreecorection}, columns are typically normalized using the $\ell_1$ norm. 

The formulation~\eqref{eq: factomat} is a constrained nonnegative matrix factorization (NMF) problem~\cite{book}, where a nonnegative matrix is approximated by a low-rank product of nonnegative matrices, possibly subject to constraints. More precisely, it is a symmetric nonnegative matrix trifactorization problem with an orthogonality constraint on the columns of $Z$. Connections between block models and NMF have been explored previously. For instance, NMF has been related to the mixed membership stochastic block model, as noted in \cite{fu2019nonnegative}, and Zhang et al.~\cite{zhang2018equivalence} approximated inference in SBM and its variants using an NMF model that relaxes the constraint that imposes non-overlapping communities.
In contrast, our formulation of the DCBM preserves the constraints without relaxation by employing an orthogonality constraint that is well-established in the NMF literature. We show that maximum-likelihood estimation under the DCBM of Karrer and Newman is equivalent to solving this constrained trifactorization problem. We analyze this connection in detail and use it to improve community detection.

For the DCBM of Karrer and Newman~\cite{SBM_degreecorection} that relies on the Poisson distribution~\eqref{eq:DCBM-Poisson}, maximizing the log-likelihood~\eqref{eq: obj_DCBM} is equivalent to minimizing the Kullback-Leibler (KL) divergence between $A$ and $B := Z\theta Z^\top$: 
\begin{equation} \label{eq:KLdiv} 
d_{\text{KL}}(A,B) = \sum_{i,j}A_{i,j}\log\left(\frac{A_{i,j}}{ B_{i,j}}\right) - A_{i,j} + B_{i,j}. 
\end{equation} 
In fact, maximizing \eqref{eq: DCBM proba} is equivalent to maximizing its logarithm, and by discarding constants, we have: 
\begin{align*}
\max_{Z,\theta} & \sum_{j<i}A_{i,j}\log\left(\left(Z\theta Z^\top\right)_{i,j}\right)-\left(Z\theta Z^\top\right)_{i,j} \notag\\
    & \quad +\sum_{i} \frac{1}{2}A_{i,i}\log\left(\left(Z\theta Z^\top\right)_{i,i}\right)-\frac{1}{2}\left(Z\theta Z^\top\right)_{i,i}.
\end{align*}
By multiplying by two and exploiting the symmetry of the problem, we finally obtain the following problem:
\begin{equation*}
\max_{Z,\theta}\sum_{i,j}A_{i,j}\log\left(\left(Z\theta Z^\top\right)_{i,j}\right)-\left(Z\theta Z^\top\right)_{i,j}, 
\end{equation*} 
which is equivalent to minimizing the KL divergence~\eqref{eq:KLdiv}.

\subsection{The OtrisymNMF model}

Instead of using the KL divergence, we propose to measure the reconstruction error in~\eqref{eq: factomat} using the squared Frobenius norm:
\begin{equation}
d_{\text{F}}( A , Z \theta Z^\top ) = \left\|A - Z\theta Z^\top \right\|_F^2 = \sum_{i,j} \left( A - Z\theta Z^\top \right)_{i,j}^2.
\label{eq: error fro}
\end{equation}
The squared Frobenius norm is widely used in matrix factorization due to its computational simplicity and favorable optimization properties. It is also implicitly used in spectral methods that are widely used for community detection. In addition, the KL divergence has intrinsic drawbacks that the Frobenius norm does not have. One limitation of the KL divergence is its reliance on a Poisson distribution, which allows for multi-edges and can lead to errors for dense graphs. Moreover, under the KL divergence, if $(Z\theta Z^\top)_{i,j}=0$, then the probability of observing a nonzero entry $A_{i,j}$ is exactly zero since we have $P(A_{i,j}=0)=1$ for a Poisson distribution of parameter $(Z\theta Z^\top)_{i,j}=0$. This implies that positive entries of $A$ cannot be approximated by zeros; that is,  $A_{i,j}>0$ implies $(Z\theta Z^\top)_{i,j}>0$; otherwise, the objective function in~\eqref{eq:KLdiv} goes to infinity. A simple example where this issue arises is when the number of communities $r$ is underestimated or when there are nodes that do not belong to any community. 

\paragraph*{Example 1} For illustration, we construct a graph consisting of two well-defined communities and one isolated node with a self-loop. Its adjacency matrix is:
\begin{equation}
    A=\begin{pmatrix}
1&1&0&0&0\\
1&1&0&0&0\\
0&0&1&1&0\\
0&0&1&1&0\\
0&0&0&0&1
    \end{pmatrix}. 
\end{equation}
Assuming $ r = 2 $, there are 15 possible non-empty partitions. Under the DCBM, the likelihood is maximized when the isolated node is assigned to one of the two communities. Due to symmetry, the KL solution is therefore not unique, there are two isolated global minima. The parameters $ Z $ and $\theta$ can be computed directly, and the expected adjacency matrix (the matrix of Poisson parameters) is: 
\begin{align}
    Z\theta Z^\top&=
    \begin{pmatrix}
        \sqrt{2}/2&0\\ \sqrt{2}/2&0\\0&2/3\\0&2/3\\0&1/3
    \end{pmatrix}\begin{pmatrix}
        2 &0\\0&9/5
    \end{pmatrix}
    \begin{pmatrix}
        \sqrt{2}/2&0\\ \sqrt{2}/2&0\\0&2/3\\0&2/3\\0&1/3
    \end{pmatrix}^\top  \notag \\
    &=\begin{pmatrix}
          1&1&0&0&0\\1&1&0&0&0\\0&0&0.8&0.8&0.4\\0&0&0.8&0.8&0.4\\0&0&0.4&0.4&0.2    
    \end{pmatrix} \quad \text{with KL,}
\end{align}
when the isolated node is assigned to the second community. We observe that the expected numbers of edges between node 5 and nodes 3 and 4 are nonzero due to the assignment of the isolated node to a community. 

In contrast, when minimizing the Frobenius norm, the optimal parameters yield:
\begin{align}
    (Z\theta Z^\top)&=
    \begin{pmatrix}
        \sqrt{2}/2&0\\\sqrt{2}/2&0\\0&\sqrt{2}/2\\0&\sqrt{2}/2\\0&0
    \end{pmatrix}\begin{pmatrix}
        2 &0\\0&2
    \end{pmatrix}
    \begin{pmatrix}
        \sqrt{2}/2&0\\\sqrt{2}/2&0\\0&\sqrt{2}/2\\0&\sqrt{2}/2\\0&0
    \end{pmatrix}^\top  \notag \\
    &=\begin{pmatrix} 
        1&1&0&0&0\\1&1&0&0&0\\0&0&1&1&0\\0&0&1&1&0\\0&0&0&0&0
    \end{pmatrix} \quad \text{with Frobenius.}
\end{align}
Here, the isolated node is not assigned to any community, which is reflected by the corresponding row of zeros in $Z$. This example highlights a limitation of the DCBM under the KL divergence: its inability to approximate positive entries by zeros may hinder its capacity to accurately capture certain structural features. On the other hand, the Frobenius norm can approximate positive entries with zeros,  making it more effective in revealing rank underestimation or structural sparsity. 

However, the Frobenius norm assigns equal weights to all reconstruction errors, regardless of their noise variance. Under the Bernoulli DCBM, the variance of each adjacency entry is given by $\operatorname{Var}(A_{ij})=P_{ij}(1-P_{ij})$, where $P_{ij}=\mathbb{E}[A_{ij}]$. Since these probabilities depend on the degree parameters of the corresponding nodes, the noise can be highly heteroscedastic in networks with heterogeneous degrees. To reduce this degree-dependent variation in noise, we follow the  degree-normalization principle used in spectral methods~\cite{score+,qin2013regularized} and approximate the regularized normalized adjacency matrix instead of the original adjacency matrix. This normalization aims to bring the noise variances to a comparable scale, while regularization improves stability for nodes with very small degrees, particularly in sparse networks. Let $D=\operatorname{diag}(d_1,\ldots,d_n)$ be the degree matrix and let $\tau=\frac{1}{n}\sum_{i=1}^n d_i$ be the average node degree. The regularized normalized adjacency matrix is defined as $L_\tau=(D+\tau I_n)^{-1/2}A(D+\tau I_n)^{-1/2}$. Minimizing its squared Frobenius reconstruction error is equivalent to minimizing a weighted Frobenius error on the original adjacency matrix:
\begin{equation}
\begin{aligned}
d_{\text{F}}(L_\tau,Z\theta Z^\top)
&=\sum_{i,j}
\left(
\frac{A_{ij}}{\sqrt{(d_i+\tau)(d_j+\tau)}}
-(Z\theta Z^\top){ij}
\right)^2\\
&=\sum_{i,j}
\frac{(A-Z'\theta Z'^\top)_{ij}^2}
{(d_i+\tau)(d_j+\tau)}\\
&=d_{\text{WF}}(A,Z'\theta Z'^\top),
\end{aligned}
\label{eq}
\end{equation}
where $Z'=(D+\tau I_n)^{1/2}Z$. This weighting reduces the contribution of reconstruction errors associated with high-degree nodes, mitigating the effect of degree-dependent noise heterogeneity.

We investigate orthogonal symmetric nonnegative matrix trifactorization (OtrisymNMF) as an alternative to the classic DCBM inference. It retains the same factorization and constraints but replaces the KL divergence with a squared Frobenius error, which offers analytical and computational advantages:
\begin{equation} 
\min_{Z \in \mathbb{R}^{n \times r}_+, 
\theta \in \mathbb{R}^{r \times r}_+} 
\left\|X - Z\theta Z^\top \right\|_F^2
\; \text{ s.t. } \; Z^\top Z = I_r,\;\theta^\top=\theta,
\label{eq: OtrisymNMF}
\end{equation}
where $X$ is a symmetric nonnegative input matrix. We use $X=L_\tau$ throughout our experiments to account for degree-dependent variation in noise. As shown in~\eqref{eq}, this corresponds to a weighted Frobenius error on the original adjacency matrix $A$. Our experiments show that this formulation, combined with the FROST algorithm introduced below, achieves community recovery comparable to, and sometimes better than, that of the DCBM inference methods considered, while being faster.
 
This model was originally introduced in~\cite{OtrisymNMFforclustering} for clustering tasks. Paul and Chen~\cite{OtrisymNMFofLapla} relate OtrisymNMF to spectral clustering through an additional nonnegativity constraint and establish the consistency of global minimizers under the SBM and DCBM. They use multiplicative updates and an additive-update variant, initialized by regularized spectral clustering. Their simulations focus on assortative networks, with spectral methods and likelihood-based refinement procedures as baselines. Here, we show that DCBM maximum-likelihood estimation and OtrisymNMF use the same factorization and constraints, but differ in the reconstruction error. This connection motivates a comparison of the two formulations and the algorithms used to solve them. We compare our method with DCBM inference algorithms in terms of community recovery and computational cost.

In our preliminary conference paper~\cite{OtrisymNMF}, we studied the OtrisymNMF model for clustering and proposed a block-coordinate descent algorithm that outperformed the multiplicative-update method of~\cite{OtrisymNMFforclustering}. The present paper extends this work to community detection under the DCBM. We present FROST, which retains the underlying block-coordinate descent scheme with a more efficient implementation for sparse adjacency matrices. We also provide a detailed explanation of the initialization strategy and its foundations in separable NMF, expanding on the brief presentation in the conference paper. Beyond its use for FROST, we extend this initialization to existing DCBM inference algorithms. Both FROST and its initialization do not assume any prior structure of the DCBM.

\subsection{FROST: An Algorithm for OtrisymNMF}

As opposed to the DCBM, for a fixed partition (that is, a fixed support of $Z$), there does not exist, to the best of our knowledge, an explicit closed-form solution for $Z$ or $\theta$ that solves \eqref{eq: OtrisymNMF}. As a result, inference strategies used for the DCBM are not applicable.  To solve the OtrisymNMF problem~\eqref{eq: OtrisymNMF}, we propose FROST (FRobenius Orthogonal Symmetric Trifactorization), which employs an alternating optimization procedure commonly used in iterative NMF methods. Starting from an initial pair ($Z$,$\theta$), FROST iteratively updates $Z$ while keeping $\theta$ fixed, then updates $\theta$ while keeping $Z$ fixed. These two steps are repeated until one of the following stopping criteria is satisfied: a maximum number of iterations, a negligible relative decrease in the reconstruction error, or a sufficiently small reconstruction error. 

\paragraph{Update of $\theta$} The update of $\theta$ has a closed-form solution. By considering the problem with $Z$ fixed and temporarily disregarding the constraints on $\theta$, we obtain the following optimality conditions:
\begin{equation}
\nabla_\theta \|X-Z \theta Z^\top\|_F^2
= 2 
Z^\top (Z \theta Z^\top - X) Z = 0.
\end{equation}
Since $Z^\top Z = I_r$, the optimal unconstrained solution is $\theta^* = Z^\top X Z$. It turns out that  $\theta^*$ automatically satisfies the nonnegativity constraint, as both $X$ and $Z$ are nonnegative, and the symmetry constraint, as $X$ is symmetric. Hence, $\theta^* = Z^\top X Z$ is the optimal solution of~\eqref{eq: OtrisymNMF} for $Z$ fixed. 

\paragraph{Update of $Z$} The update of $Z$ is performed using block-coordinate descent, updating one row of $Z$ at a time. In practical terms, the affiliation of one node is updated at each row update, while the affiliations of the other nodes remain fixed. For each node $i$, we assign it to the community that yields the largest decrease in reconstruction error. To do so, for each community $k \in \{ 1, \ldots, r \}$, we compute the optimal value of $Z(i,k)$ assuming that node $i$ belongs to community $k$. This amounts to solving the following one-variable subproblem in the variable $Z(i,k)$, obtained by expanding the Frobenius norm:
\begin{equation}
\begin{array}{rl}
\displaystyle \min_{Z(i,k)\geq 0}  & \left(X(i,i) - Z(i,k)\, \theta(k,k)\, Z(i,k)\right)^2 \\[1ex]
+ & 2\displaystyle\sum_{j \neq i} \left(X(i,j) - Z(i,k)\, \theta(k,:)\, Z(j,:)^\top\right)^2 + \text{const}.
\end{array}
\label{subproblem1var}
\end{equation}
Solving \eqref{subproblem1var} reduces to minimizing a univariate fourth-order polynomial of the form
\[
az^4 + bz^2 + cz,
\]
where the coefficients are given by:
\begin{itemize}
    \item $a = \theta(k,k)^2$,
    \item $b = 2\left( \sum_{j \neq i} (Z(j,:) \theta(:,k))^2 - \theta(k,k) X(i,i) \right)$,
    \item $c = -4 \sum_{j \neq i} X(i,j)\, Z(j,:) \theta(:,k)$.
\end{itemize}
We compute the stationary points using Cardano's formula and select the point with the smallest objective value among the nonnegative stationary points and the boundary point $z=0$. When the graph has no self-loops, $X(i,i)=0$, yielding $a,b\geq0$ and $c\leq0$. The objective is therefore convex, and for $a>0$, its minimizer is the unique nonnegative root of $4az^3+2bz+c=0$. For $a=0$, we set $z=-c/(2b)$ if $b>0$, and $z=0$ otherwise. Once the coefficients are available, the subproblem is solved in $\mathcal{O}(1)$ operations. If a complete partition is required, nodes corresponding to zero rows of the final $Z$ are assigned uniformly at random to one of the $r$ communities, without modifying the factorization.
Finally, after testing the $r$ communities, we update the $i$th row of $Z$ according to the community assignment that yields the largest reduction in error. Once the community assignments for all $n$ nodes have been updated, we normalize the columns of $Z$. The matrix $\theta$ is then updated accordingly.

In our implementation, the matrix $Z$ is represented using two vectors, $w$ and $v$, of length $n$. The $i$th element of $v$, $v(i) \in \{1, \ldots, r\}$,  stores the index of the community of node $i$; that is, the column index of the nonzero entry in the $i$th row of $Z$, while $w(i) = Z(i, v(i))$ stores its value. This representation gives each iteration of the algorithm a cost of $\mathcal{O}(nr\langle d \rangle)$ operations. Since $\langle d \rangle$ is the average degree, $n\langle d \rangle$ is twice the number of edges in the graph. Therefore, the algorithm scales linearly with the product of the number of edges in the graph and the number of communities.

\subsection{Robust initialization via separable NMF}

We propose an efficient initialization method for $Z$ and $\theta$, inspired by the observation that the model~\eqref{eq: factomat} exhibits a so-called separability property.

In the noiseless case, $X = W Z^\top$ with $W=Z\theta \geq 0$ and $Z \geq 0$ corresponds to a classical NMF~\cite{lee1999learning}, that is, $X$ is factorized as the product of two smaller nonnegative matrices, $W$ and $Z$. Moreover, $Z$ has additional structure, as it is orthogonal. In particular, it is separable; that is, there exists an index set $K \subset \{1,2,\dots,n\}$ of cardinality $r$ such that  $Z(K, :) = \text{diag}(z) \in \mathbb{R}^{r \times r}$, with a strictly positive vector $z \in \mathbb{R}_+^r$, where $\text{diag}(\cdot)$ denotes the diagonal matrix with $z$ on its diagonal. Equivalently, it requires that $Z$ contains, up to permutation and scaling, the identity matrix as a submatrix. In simpler terms, $Z$ is separable if for each column of $Z$, there exists at least one row with a single nonzero entry in that column.  
In summary, we have 
\begin{equation}
X(:, K) = W Z(K, :)^\top = W \, \text{diag}(z). 
\end{equation}
This means that each column of $W$ is equal to a column of $X$ up to scaling factors. Geometrically, there exists a subset of $r$ columns of $X$, $X(:,K)$, such that the convex cone they generate contains all the columns of $X$, and thus spans the entire cone generated by $X$: 
\begin{align}
    \text{cone}(X)=\text{cone}(X(:,K))=\text{cone}(W), \notag\\
    \text{where}\quad \text{cone}(X)=\{x \ | \ x=Xy, y\in\mathbb{R}_+^n \}.
\end{align}
For an illustration, see Fig.~\ref{fig:cone}.
\begin{figure}[]
  \centering
  \begin{tikzpicture}
\begin{axis}[%
    width=8cm, height=8cm,
    xmin=0, xmax=1,
    ymin=0, ymax=1,
    zmin=0, zmax=1,
    xtick={0,0.5,1},
    ytick={0,0.5,1},
    ztick={0,0.5,1},
    view={120}{35},
    axis background/.style={fill=white},
    grid=both,
    legend style={
        at={(1.2,0.95)},   
        anchor=south east,  
    },
    legend image post style={only marks},
]
\addplot3[only marks, mark=x, mark options={red, scale=2}] coordinates {(1,0.4,0)};
\addlegendentry{Columns of $W$}

\addplot3[only marks, mark=o, mark options={blue, scale=1.5}] coordinates {(0.35,0.25,0.15)};
\addlegendentry{Columns of $X$}

\addplot3[thick, blue] coordinates {(0,0,0) (1,0.4,0)};
\addplot3[only marks,  mark=o, mark options={blue, scale=1.5}] coordinates {(1,0.4,0)};
\addplot3[only marks, mark=x, mark options={red, scale=3}] coordinates {(1,0.4,0)};
\addplot3[thick, blue] coordinates {(0,0,0) (0.1,0.95,0)};
\addplot3[only marks, mark=o, mark options={blue, scale=1.5}] coordinates {(0.1,0.95,0)};
\addplot3[only marks, mark=x, mark options={red, scale=3}] coordinates {(0.1,0.95,0)};
\addplot3[ thick, blue] coordinates {(0,0,0) (0.1,0.2,1)};
\addplot3[only marks, mark=x, mark options={red, scale=3}] coordinates {(0.1,0.2,1)};
\addplot3[only marks, mark=o, mark options={blue, scale=1.5}] coordinates {(0.1,0.2,1)};

\addplot3[
    fill=red!20, 
    opacity=0.3, 
    draw=red
] coordinates {
    (0,0,0)
    (1,0.4,0)
    (0.1,0.95,0)
};

\addplot3[
    fill=red!20, 
    opacity=0.3, 
    draw=red
] coordinates {
    (0,0,0)
    (0.1,0.95,0)
    (0.1,0.2,1)
};

\addplot3[
    fill=red!20, 
    opacity=0.3, 
    draw=red
] coordinates {
    (0,0,0)
    (0.1,0.2,1)
    (1,0.4,0)
};

\foreach \x/\y/\z in {0.9/0.80/0.1, 0.77/1/0.65, 0.33/0.54/1, 
 0.58/1/0.25, 0.64/0.88/1, 1/0.57/0.36, 0.46/1/0.73, 
 0.29/0.71/0.1, 
 0.55/0.50/0.25,
  0.65/0.55/0.40,
  0.35/0.75/0.25,
  0.20/0.65/0.55,
  0.45/0.55/0.60,
  0.30/0.40/0.75,
  0.40/0.60/0.50,
  0.25/0.45/0.25,
  0.35/0.35/1} {
    \addplot3[thick, blue] coordinates {(0,0,0) (\x,\y,\z)};
    \addplot3[only marks, mark=o, mark options={blue, scale=1.5}] coordinates {(\x,\y,\z)};
}

\end{axis}
\end{tikzpicture}
  \caption{Geometric illustration of the separability property with $n=20$ and $r=3$. The extreme rays of $\text{cone}(X)$ are present in $X$ as columns and correspond to the columns of $W$ (up to scaling). For visualization, the example is represented in 3D, although the original data are $n=20$ dimensional. }
  \label{fig:cone}
\end{figure}
Finding this set $K$ of columns of $X$, which is equivalent to identifying the extreme rays of $\text{cone}(X)$, is solvable in polynomial time with provably robust algorithms in the presence of noise. In a nutshell, given $X = WZ^\top + N$ where $Z$ is separable and $N$ is the noise, $W$ can be recovered up to the noise level when it is sufficiently small. This line of research was initiated by Arora et al.~\cite{arora2012computing}; see~\cite[Chapter 7]{book} for a comprehensive account of separable NMF algorithms and their robustness to noise. 

For our application, assuming that $Z$ is separable amounts to assuming that there exists at least one node belonging to each community. In practice, there are typically multiple nodes associated with each community. Considering the definition of separability, in our case, $Z$ actually contains several disjoint diagonal submatrices of size $r$, up to permutations. This is a more favorable situation since there is more than one column of $X$ close to each column of $W$. This was recently exploited by Bhattacharyya et al.~\cite{bhattacharyya2020finding, bhattacharyya2021finding, bhattacharyya2021learning}, who use more than one column of $X$ to approximate the columns of $W$, leading to provably more robust algorithms. 
To estimate $W$ under this stronger assumption, we use the smoothed vertex component analysis method (SVCA) proposed by Nadisic et al.~\cite{nadisic2023smoothed}, a smoothed version of vertex component analysis (VCA)~\cite{VCA}, which is a widely used separable NMF algorithm. SVCA improves robustness by selecting $s$ columns of $X$ to estimate each column of $W$.  In our context, ideally, we should estimate each column of $W$ using $s$ columns of $X$, where $s$ is the number of nodes in the corresponding community. As these values are typically unknown, we choose  $s = \max\left(2, \lfloor 0.1  \frac{n}{r} \rfloor\right)$, assuming that there are at least $\max\left(2, \lfloor 0.1  \frac{n}{r} \rfloor\right)$ nodes in each community. SVCA is preferred over the method proposed by~\cite{bhattacharyya2020finding} and over the smoothed version of the successive projection algorithm (SSPA)~\cite{nadisic2023smoothed} due to its greater robustness to non-Gaussian noise, including Poisson noise, and its superior overall performance~\cite{nadisic2023smoothed}. In short, SVCA is a greedy algorithm that computes each column of $W$ sequentially by averaging over $s$ columns of $X$ selected to maximize their projection onto a particular randomly sampled direction. This direction is drawn from the subspace spanned by the top $r$ singular vectors of $X$ and further projected to be orthogonal to the columns already selected. It has a computational cost of  $\mathcal{O}(n r\langle d \rangle  )$ operations. SVCA enjoys a probabilistic robustness guarantee, ensuring that it recovers with a certain probability a matrix $W'$ such that $\|W - W' \Pi \|_F \leq \epsilon$, where $\Pi$ is a permutation matrix and $\epsilon$ depends on the noise level, the conditioning 
of $W$, and the number of points close to each column of 
$W$~\cite{nadisic2023smoothed}. 

After determining $W$ using SVCA, we recover $Z$ by solving the following problem:
\begin{equation}
\min_{Z\geq0} \| X - WZ^\top \|_F^2\quad \text{such that}\quad (Z^\top Z)_{i,j}=0\ \forall i\neq j,\label{eq:ONMFinZ}
\end{equation} 
where $Z$ can be normalized afterward due to the scaling degree of freedom. Problem~\eqref{eq:ONMFinZ} is related to orthogonal NMF (ONMF)~\cite{pompili2014two}, 
and $Z$ can be computed in closed form, as in~\cite{pompili2014two}. This  amounts to assigning each column of $X$ to its closest cluster centroid given by the columns of $W$, where the closeness is measured in terms of angles. 

For FROST, this procedure based on SVCA initializes both $Z$ and $\theta = Z^\top X Z$. The method is extremely fast, perfectly recovers $Z$ and $\theta$ in the noiseless case, and is robust to noise, not limited to Gaussian noise. For DCBM inference, the SVCA initialization can also be used  to provide an initial node partition derived from the computed matrix $Z$, which serves as the starting point for the inference methods. As SVCA is a randomized algorithm, it can be run multiple times and the initialization yielding the lowest error can be selected. In our experiments, however, when used as an initialization method, we perform a single run of SVCA per initialization.

\paragraph*{Related work} 

Separable NMF-based methods have already been used for community detection under mixed membership SBMs, which extend the SBM to allow overlapping communities. To ensure model separability, they assume that each community contains at least one pure node, a node that belongs to only one community. Under this assumption, separable NMF methods can identify one pure node per community, facilitating the estimation of mixed memberships and other model parameters. For instance, Jin et al.~\cite{jin2017estimating} proposed a spectral method for estimating mixed membership in a degree-corrected mixed-membership model with a step to identify pure nodes. Similarly, Panov et al.~\cite{panov2017consistent} studied consistent estimation for the  mixed membership SBM under the same assumption of separability. More recently, Qing~\cite{qingmultilayer} proposed a spectral method based on separable NMF to estimate the overlapping communities in multi-layer directed networks. For topic modeling, Arora et al.~\cite{TrisymNMF_separable} presented a symmetric nonnegative matrix trifactorization model without orthogonality constraints, allowing words to belong to multiple topics. The model is based on the separability assumption, meaning that each topic contains at least one word used exclusively by that topic. Bhattacharyya et al.~\cite{bhattacharyya2020finding} initiated the concept of smoothed separable NMF and proposed using their smoothed separable NMF method to estimate the parameters of mixed-membership SBMs, assuming that each community contains multiple pure nodes. This is a primarily theoretical work and does not provide empirical comparisons with other methods.

In our setting with disjoint communities, the assumption that multiple columns of $X$ are close to each column of $W$ is even more strongly justified, as all nodes belong to a single community. Moreover, the parameters of our model, $Z$ and $\theta$, are determined through a closed-form expression. To the best of our knowledge, this is the first time such properties have been leveraged for the DCBM.

\section{Numerical experiments}

We compare OtrisymNMF and the DCBM of Karrer and Newman~\cite{SBM_degreecorection}, together with their respective optimization algorithms, on synthetic and real networks. We additionally include graph-tool~\cite{graph-tool} and SCORE+~\cite{score+} as baselines (see below) and evaluate the effectiveness of our SVCA initialization strategy for DCBM inference. To compare the community assignments produced by the methods with the true assignments, we use the asymmetrically normalized reduced mutual information (RMI)~\cite{RMI}. Conventional NMI can favor partitions with many small communities and its symmetric normalization introduces additional bias. RMI addresses these limitations. RMI equals $1$ for identical partitions and can take negative values for uninformative partitions.

For the Poisson DCBM, we consider two algorithms that optimize~\eqref{eq: obj_DCBM}. The first, denoted by \textbf{KN}, is the algorithm of Karrer and Newman~\cite{SBM_degreecorection}, inspired by the Kernighan--Lin algorithm~\cite{Kernighan}.  During each iteration, nodes are moved sequentially to the community that either increases the objective function the most or decreases it the least. The partition with the best objective value is retained, and iterations continue until no further improvement is obtained. The second, \textbf{KL-EM}~\cite{SBMcode}, evaluates candidate node moves from the same partition and applies the best updates simultaneously. Each iteration of either method costs $\mathcal{O}\left(nr(\min(r,\langle d\rangle)+\langle d\rangle)\right)$, where $\langle d\rangle$ is the average degree. We also include \textbf{graph-tool}~\cite{graph-tool}, using Peixoto's multilevel agglomerative algorithm~\cite{peixoto2014efficient}. It performs Bayesian inference for a degree-corrected block model, incorporating priors to balance model fit and complexity. Its objective therefore differs from the Poisson DCBM log-likelihood in~\eqref{eq: obj_DCBM}. We fix the number of communities to $r$. For sparse networks, the method has a reported complexity of $\mathcal{O}(n\log^2 n)$. For KN and KL-EM, we use the implementation provided in~\cite{SBMcode}, and for graph-tool, we use the official implementation~\cite{graph-tool}. For runs initialized with SVCA, FROST, KN, and KL-EM receive the same initial partition, whereas graph-tool uses its own agglomerative initialization. Finally, we include \textbf{SCORE+}~\cite{score+} as an efficient spectral baseline designed to handle degree heterogeneity and weak community signals under the DCBM. We limit the spectral comparison to this method, as our main focus is on matrix factorization and DCBM inference.

All experiments were performed using Python 3.12 on a laptop with 2.80~GHz Intel\textsuperscript{\textregistered} Core\texttrademark{} i7-1165G7. In this paper, runtime refers to the elapsed time. All experiments can be reproduced using the code available on GitHub~\url{https://github.com/Alexia1305/OtrisymNMF_DCBM}.

\subsection{DCBM Synthetic Networks}

We compare FROST with DCBM inference methods (KN, KL-EM, and graph-tool) and evaluate the SVCA initialization on synthetic graphs generated from the Bernoulli DCBM~\eqref{eq: Bernoulli DCBM}. Unless otherwise specified, we use $n=600$ nodes, $r=3$ equally sized communities, and an expected average degree of $\alpha=10$. To introduce degree heterogeneity, the nonzero values of $Z$ are independently drawn from a power-law (Pareto) distribution (as in~\cite{score+,qin2013regularized}) with lower bound $x_{\min}=1$ and exponent $\gamma=2.5$. The probability density function is $p(x)=(\gamma-1)x^{-\gamma}$, for $x\geq 1$.
We define $\beta=p/q$, where $p$ and $q$ are the diagonal and off-diagonal entries of the matrix $\theta$, respectively. By default, we set $\beta=6$, corresponding to approximately three times as many within-community edges as between-community edges for $r=3$. For each synthetic network, we run FROST, KN, KL-EM, and graph-tool 10 times and retain, for each method, the solution with the best value of its respective objective function. The only exception is the scalability experiments, where each method is run only once per network.

\paragraph{Community recovery across assortative and disassortative regimes}

\begin{figure}[t]
    \centering
    \begin{tikzpicture}
        \begin{axis}[
            width=0.9\linewidth,
            height=7cm,
            xmode=log,
            xmin=0.045,
            xmax=22,
            ymin=0,
            ymax=1,
            xlabel={$\beta$},
            ylabel={RMI},
            xtick={0.05,0.1,0.25,0.5,1,2,4,10,20},
            xticklabels={$0.05$,$0.1$,$0.25$,$0.5$,$1$,$2$,$4$,$10$,$20$},
            grid=major,
            major grid style={gray!20},
            legend style={
                at={(0.35,0.97)},
                anchor=north,
                legend columns=2,
                font=\small,
                /tikz/column 2/.style={column sep=1em}
            },
            tick label style={font=\small},
            label style={font=\small},
            every axis plot/.append style={thick, mark size=1.8pt},
        ]

    \addplot+[blue, solid, mark=*, mark options={fill=blue}] coordinates {
        (0.05,0.3228) (0.1,0.1703) (0.25,0.0225) (0.5,0.0010) (0.75,0.0004)
        (1,0.0004) (1.5,0.0005) (2,0.0021) (2.5,0.0152) (3,0.0526)
        (4,0.2430) (6,0.5101) (10,0.6633) (20,0.8575)
    };
    \addlegendentry{KN}
    
    \addplot+[red, solid, mark=*, mark options={fill=red}] coordinates {
        (0.05,0.5749) (0.1,0.4261) (0.25,0.0632) (0.5,0.0025) (0.75,0.0003)
        (1,0.0002) (1.5,0.0008) (2,0.0039) (2.5,0.0227) (3,0.1306)
        (4,0.4521) (6,0.7013) (10,0.8363) (20,0.9275)
    };
    \addlegendentry{KL-EM}
    
    \addplot+[blue, dashed, mark=square*, mark options={fill=blue,solid}] coordinates {
        (0.05,0.5954) (0.1,0.4907) (0.25,0.1972) (0.5,0.0062) (0.75,0.0008)
        (1,0.0003) (1.5,0.0011) (2,0.0140) (2.5,0.1235) (3,0.2757)
        (4,0.4910) (6,0.7018) (10,0.8367) (20,0.9265)
    };
    \addlegendentry{KN(S)}
    
    \addplot+[red, dashed, mark=square*, mark options={fill=red,solid}] coordinates {
        (0.05,0.6037) (0.1,0.5006) (0.25,0.2128) (0.5,0.0064) (0.75,0.0006)
        (1,0.0002) (1.5,0.0007) (2,0.0137) (2.5,0.1210) (3,0.2826)
        (4,0.4960) (6,0.6996) (10,0.8386) (20,0.9276)
    };
    \addlegendentry{KL-EM(S)}
    
    \addplot+[darkgreen, solid, mark=x] coordinates {
        (0.05,0.6108) (0.1,0.4239) (0.25,0.0002) (0.5,0.0001) (0.75,0.0003)
        (1,0.0004) (1.5,0.0025) (2,0.0028) (2.5,0.0042) (3,0.0100)
        (4,0.3539) (6,0.6945) (10,0.8407) (20,0.9317)
    };
    \addlegendentry{graph-tool}
    
    \addplot+[cyan, solid, mark=star] coordinates {
        (0.05,0.3689) (0.1,0.3028) (0.25,0.1521) (0.5,0.0173) (0.75,0.0014)
        (1,0.0005) (1.5,0.0013) (2,0.0242) (2.5,0.1393) (3,0.2691)
        (4,0.4571) (6,0.6464) (10,0.7945) (20,0.8899)
    };
    \addlegendentry{SCORE+}
    
    \addplot+[black, dashed, mark=triangle*, mark options={fill=black,solid}] coordinates {
        (0.05,0.6107) (0.1,0.5163) (0.25,0.2279) (0.5,0.0083) (0.75,0.0004)
        (1,0.0001) (1.5,0.0007) (2,0.0203) (2.5,0.1473) (3,0.3010)
        (4,0.4967) (6,0.6867) (10,0.8189) (20,0.9141)
    };
    \addlegendentry{FROST(S)}

        \addplot[dashed, black, forget plot] coordinates {(1,0) (1,1)};

        \end{axis}
    \end{tikzpicture}
    \caption{Average RMI over 100 sampled networks as a function of $\beta$. The vertical line at $\beta=1$ separates the disassortative and assortative regimes. (S) indicates SVCA initialization.}
    \label{fig:beta}
\end{figure}

We first vary $\beta$ from $0.05$ to $20$ to examine community recovery in disassortative ($\beta<1$) and assortative ($\beta>1$) graphs. 
Figure~\ref{fig:beta} shows the average RMI over 100 independently sampled networks. FROST achieves the highest RMI in several challenging regimes, particularly for disassortative graphs, while likelihood-based methods perform slightly better under strong assortativity. Graph-tool performs poorly at intermediate levels of community separation, where FROST still recovers part of the structure. SVCA initialization substantially improves KN and KL-EM. SCORE+ is competitive in the assortative regime, with scores generally slightly below those of FROST, but struggles to recover communities in the disassortative regime.

\begin{figure}[t]
    \centering
    \begin{tikzpicture}
        \begin{groupplot}[
            group style={
                group size=2 by 1,
                horizontal sep=0.03\linewidth,
            },
            scale only axis,
            width=0.425\linewidth,
            height=5.3cm,
            ymin=0, ymax=1,
            ytick={0,0.2,0.4,0.6,0.8,1},
            tick label style={font=\small},
            label style={font=\small},
            every axis plot/.append style={thick, mark size=1.5pt},
            grid=major,
            major grid style={gray!20},
        ]

        \nextgroupplot[
            xmode=log,
            xmin=4.5, xmax=23,
            xlabel={Average degree},
            ylabel={RMI},
            xtick={5,6,8,10,15,20},
            xticklabels={$5$,$6$,$8$,$10$,$15$,$20$},
        ]

        \addplot+[blue, solid, mark=*, mark options={fill=blue}] coordinates {
            (5,0.1239) (6,0.2196) (8,0.3779) (10,0.5101) (15,0.7112) (20,0.8091)
        };
        
        \addplot+[red, solid, mark=*, mark options={fill=red}] coordinates {
            (5,0.3155) (6,0.4300) (8,0.5934) (10,0.7013) (15,0.8420) (20,0.9141)
        };
        
        \addplot+[blue, dashed, mark=square*, mark options={fill=blue,solid}] coordinates {
            (5,0.4229) (6,0.4910) (8,0.6066) (10,0.7018) (15,0.8415) (20,0.9138)
        };
        
        \addplot+[red, dashed, mark=square*, mark options={fill=red,solid}] coordinates {
            (5,0.4270) (6,0.4951) (8,0.6034) (10,0.6996) (15,0.8375) (20,0.9141)
        };
        
        \addplot+[darkgreen, solid, mark=x] coordinates {
            (5,0.1671) (6,0.3651) (8,0.5731) (10,0.6945) (15,0.8403) (20,0.9145)
        };
        
        \addplot+[cyan, solid, mark=star] coordinates {
            (5,0.2919) (6,0.4413) (8,0.5561) (10,0.6464) (15,0.7897) (20,0.8645)
        };
        
        \addplot+[black, dashed, mark=triangle*, mark options={fill=black,solid}] coordinates {
            (5,0.4360) (6,0.4954) (8,0.5978) (10,0.6867) (15,0.8201) (20,0.8943)
        };

        \nextgroupplot[
            xmin=1.9, xmax=3.1,
            xlabel={$\gamma$},
            yticklabels=\empty,
            xtick={2,2.5,3},
            grid=major,
            major grid style={gray!20},
            legend style={
                at={(0.97,0.03)},
                anchor=south east,
                legend columns=2,
                font=\small,
                row sep=-1pt,
                inner sep=1.5pt,
                cells={anchor=west},
                /tikz/column 2/.style={column sep=0.5em},
            },
        ]

        \addplot+[blue, solid, mark=*, mark options={fill=blue}] coordinates {
            (2,0.3270) (2.25,0.4117) (2.5,0.5101) (2.75,0.5334) (3,0.5903) 
        };
        \addlegendentry{KN}
        
        \addplot+[red, solid, mark=*, mark options={fill=red}] coordinates {
            (2,0.4206) (2.25,0.5975) (2.5,0.7013) (2.75,0.7394) (3,0.7731) 
        };
        \addlegendentry{KL-EM}
        
        \addplot+[blue, dashed, mark=square*, mark options={fill=blue,solid}] coordinates {
            (2,0.4458) (2.25,0.6153) (2.5,0.7018) (2.75,0.7400) (3,0.7757) 
        };
        \addlegendentry{KN(S)}
        
        \addplot+[red, dashed, mark=square*, mark options={fill=red,solid}] coordinates {
            (2,0.4510) (2.25,0.6105) (2.5,0.6996) (2.75,0.7431) (3,0.7769) 
        };
        \addlegendentry{KL-EM(S)}
        
        \addplot+[darkgreen, solid, mark=x] coordinates {
            (2,0.3072) (2.25,0.5648) (2.5,0.6945) (2.75,0.7438) (3,0.7724) 
        };
        \addlegendentry{graph-tool}
        
        \addplot+[cyan, solid, mark=star] coordinates {
            (2,0.3768) (2.25,0.5462) (2.5,0.6464) (2.75,0.6956) (3,0.7268) 
        };
        \addlegendentry{SCORE+}
        
        \addplot+[black, dashed, mark=triangle*, mark options={fill=black,solid}] coordinates {
            (2,0.4566) (2.25,0.5987) (2.5,0.6867) (2.75,0.7264) (3,0.7553) 
        };
        \addlegendentry{FROST(S)}

        \end{groupplot}
    \end{tikzpicture}
    \caption{Average RMI over 100 sampled networks as a function of the average degree
    (left) and the power-law exponent $\gamma$
    (right).}
    \label{fig:sparsity_heterogeneity}
\end{figure}
\paragraph{Sparsity}
This experiment compares the methods across different levels of graph sparsity by varying the average degree from $5$ to $20$. The left panel of Figure~\ref{fig:sparsity_heterogeneity} shows the average RMI as a function of the average degree. At high average degrees, graph-tool performs slightly better than FROST, KN(S), and KL-EM. SVCA initialization substantially improves KN across the tested range and improves KL-EM on sparse graphs. Below an average degree of $10$, FROST performs best or on par with KN(S) and KL-EM(S), while graph-tool's performance deteriorates sharply.

\paragraph{Degree heterogeneity}
This experiment examines how degree heterogeneity affects the performance of the methods. We vary $\gamma$ from $2$ to $3$. A smaller $\gamma$ indicates greater degree heterogeneity. The right panel of Figure~\ref{fig:sparsity_heterogeneity} shows the average RMI as a function of $\gamma$. FROST, as well as KN and KL-EM initialized with SVCA, achieve the highest RMI under strong degree heterogeneity ($\gamma<2.5$), while graph-tool's performance deteriorates as $\gamma$ decreases. SVCA initialization again substantially improves KN across the tested range.

\begin{figure*}[t]
    \centering
    \begin{tikzpicture}

    \begin{groupplot}[
        group style={
            group size=2 by 2,
            horizontal sep=0.9cm,
            vertical sep=1.1cm,
        },
        width=0.52\linewidth,
        height=7cm,
        xmode=log,
        grid=major,
        major grid style={gray!20},
        tick label style={font=\small},
        label style={font=\small},
        title style={font=\small},
        every axis plot/.append style={
            thick,
            mark size=1.8pt,
        },
    ]

    \nextgroupplot[
        xlabel={$n$},
        ylabel={Time (s)},
        ymode=log,
        xmin=2500,
        xmax=120000,
        xtick={3000,6000,10000,20000,30000,60000,100000},
        xticklabels={
            3k,6k,10k,20k,30k,60k,100k
        }
    ]

    \addplot+[blue, solid, mark=*, mark options={fill=blue}] coordinates {
        (3000,1.7080) (6000,2.9675) (10000,8.3274) (20000,17.3699) (30000,30.6048) (60000,82.9565) (100000,207.7807)
    };
    
    \addplot+[red, solid, mark=*, mark options={fill=red}] coordinates {
        (3000,3.1182) (6000,6.2288) (10000,11.1871) (20000,26.5948) (30000,46.6186) (60000,88.1097) (100000,147.7398)
    };
    
    \addplot+[blue, dashed, mark=square*, mark options={fill=blue,solid}] coordinates {
        (3000,1.3497) (6000,2.4181) (10000,5.4903) (20000,12.6958) (30000,18.5888) (60000,49.9093) (100000,107.5484)
    };
    
    \addplot+[red, dashed, mark=square*, mark options={fill=red,solid}] coordinates {
        (3000,1.7535) (6000,3.1675) (10000,8.2516) (20000,18.8502) (30000,31.6284) (60000,70.8931) (100000,116.7887)
    };
    
    \addplot+[darkgreen, solid, mark=x] coordinates {
        (3000,5.0430) (6000,14.1692) (10000,24.6942) (20000,66.9635) (30000,129.9277) (60000,401.9766) (100000,535.6759)
    };
    
    \addplot+[cyan, solid, mark=star] coordinates {
        (3000,0.2239) (6000,0.3009) (10000,0.3256) (20000,4.3400) (30000,4.3621) (60000,6.6167) (100000,11.4307)
    };
    
    \addplot+[black, dashed, mark=triangle*, mark options={fill=black,solid}] coordinates {
        (3000,0.7595) (6000,1.3831) (10000,2.6620) (20000,5.5317) (30000,8.6922) (60000,18.9898) (100000,32.6916)
    };

    \nextgroupplot[
        xlabel={$r$},
        ymode=log,
        xmin=2.5,
        xmax=60,
        xtick={3,5,10,20,30,50},
        xticklabels={$3$,$5$,$10$,$20$,$30$,$50$},
    ]

    \addplot+[blue, solid, mark=*, mark options={fill=blue}] coordinates {
        (3,31.9958) (5,65.1306) (10,149.5437) (20,296.4916) (30,537.6398) (50,868.7621)
    };
    
    \addplot+[red, solid, mark=*, mark options={fill=red}] coordinates {
        (3,47.8140) (5,70.9364) (10,178.7893) (20,375.8394) (30,622.9566) (50,1359.0913)
    };
    
    \addplot+[blue, dashed, mark=square*, mark options={fill=blue,solid}] coordinates {
        (3,21.2031) (5,34.5840) (10,61.4224) (20,157.3743) (30,324.1144) (50,663.2639)
    };
    
    \addplot+[red, dashed, mark=square*, mark options={fill=red,solid}] coordinates {
        (3,33.3142) (5,58.1775) (10,82.2524) (20,263.4209) (30,471.2506) (50,971.1435)
    };
    
    \addplot+[darkgreen, solid, mark=x] coordinates {
        (3,142.7684) (5,118.1156) (10,159.5278) (20,91.7271) (30,86.0787) (50,77.8276)
    };
    
    \addplot+[cyan, solid, mark=star] coordinates {
        (3,4.9986) (5,5.3504) (10,6.9595) (20,9.0505) (30,16.9350) (50,25.7682)
    };
    
    \addplot+[black, dashed, mark=triangle*, mark options={fill=black,solid}] coordinates {
        (3,9.1402) (5,9.8445) (10,11.7668) (20,15.5979) (30,23.1082) (50,43.8360)
    };

    \nextgroupplot[
        xlabel={$n$},
        ylabel={RMI},
        xmin=2500,
        xmax=120000,
        ymin=0,
        ymax=1,
        ytick={0,0.2,0.4,0.6,0.8,1},
        xtick={3000,6000,10000,20000,30000,60000,100000},
        xticklabels={
            3k,6k,10k,20k,30k,60k,100k
        },
    ]

    \addplot+[blue, solid, mark=*, mark options={fill=blue}] coordinates {
        (3000,0.1968) (6000,0.1832) (10000,0.2771) (20000,0.2109) (30000,0.1387) (60000,0.0399) (100000,0.1075)
    };
    
    \addplot+[red, solid, mark=*, mark options={fill=red}] coordinates {
        (3000,0.5336) (6000,0.4694) (10000,0.3993) (20000,0.5608) (30000,0.5311) (60000,0.1875) (100000,0.2045)
    };
    
    \addplot+[blue, dashed, mark=square*, mark options={fill=blue,solid}] coordinates {
        (3000,0.6016) (6000,0.6747) (10000,0.6676) (20000,0.6345) (30000,0.6279) (60000,0.6609) (100000,0.6552)
    };
    
    \addplot+[red, dashed, mark=square*, mark options={fill=red,solid}] coordinates {
        (3000,0.6091) (6000,0.6782) (10000,0.6719) (20000,0.6392) (30000,0.6395) (60000,0.6670) (100000,0.6625)
    };
    
    \addplot+[darkgreen, solid, mark=x] coordinates {
        (3000,0.6633) (6000,0.6687) (10000,0.6599) (20000,0.6581) (30000,0.6478) (60000,0.5764) (100000,0.4210)
    };
    
    \addplot+[cyan, solid, mark=star] coordinates {
        (3000,0.5672) (6000,0.4786) (10000,0.4383) (20000,0.4180) (30000,0.3521) (60000,0.3924) (100000,0.2772)
    };
    
    \addplot+[black, dashed, mark=triangle*, mark options={fill=black,solid}] coordinates {
        (3000,0.5894) (6000,0.6552) (10000,0.6185) (20000,0.5819) (30000,0.6080) (60000,0.6376) (100000,0.6316)
    };

    \nextgroupplot[
        xlabel={$r$},
        xmin=2.5,
        xmax=60,
        ymin=0,
        ymax=1,
        ytick={0,0.2,0.4,0.6,0.8,1},
        xtick={3,5,10,20,30,50},
        xticklabels={$3$,$5$,$10$,$20$,$30$,$50$},
        legend style={
            at={(0.245,0.37)},
            anchor=north west,
            draw=none,
            font=\small,
            legend columns=2,
        },
    ]

        \addplot+[blue, solid, mark=*, mark options={fill=blue}] coordinates {
            (3,0.1387) (5,0.3389) (10,0.5874) (20,0.6903) (30,0.7497) (50,0.7887)
        };
        \addlegendentry{KN}
        
        \addplot+[red, solid, mark=*, mark options={fill=red}] coordinates {
            (3,0.5311) (5,0.7330) (10,0.7683) (20,0.8037) (30,0.8307) (50,0.8386)
        };
        \addlegendentry{KL-EM}
        
        \addplot+[blue, dashed, mark=square*, mark options={fill=blue,solid}] coordinates {
            (3,0.6279) (5,0.7491) (10,0.8188) (20,0.8287) (30,0.8511) (50,0.8577)
        };
        \addlegendentry{KN(S)}
        
        \addplot+[red, dashed, mark=square*, mark options={fill=red,solid}] coordinates {
            (3,0.6395) (5,0.7658) (10,0.8199) (20,0.8352) (30,0.8547) (50,0.8577)
        };
        \addlegendentry{KL-EM(S)}
        
        \addplot+[darkgreen, solid, mark=x] coordinates {
            (3,0.6478) (5,0.7788) (10,0.8507) (20,0.8763) (30,0.8862) (50,0.8977)
        };
        \addlegendentry{graph-tool}
        
        \addplot+[cyan, solid, mark=star] coordinates {
            (3,0.3521) (5,0.5652) (10,0.5993) (20,0.6617) (30,0.7477) (50,0.7767)
        };
        \addlegendentry{SCORE+}
        
        \addplot+[black, dashed, mark=triangle*, mark options={fill=black,solid}] coordinates {
            (3,0.6080) (5,0.6936) (10,0.7866) (20,0.7787) (30,0.8189) (50,0.8139)
        };
        \addlegendentry{FROST(S)}

    \end{groupplot}
    \end{tikzpicture}

    \caption{
        Scalability with respect to the number of nodes $n$ (left)
        and communities $r$ (right): average running time (top)
        and RMI (bottom) over 10 independently generated networks
        per configuration.
    }
    \label{fig:scalability}
\end{figure*}

\paragraph{Scalability}

We assess the scalability of the methods by varying the number of nodes and the number of communities. First, we vary $n$ from $3{,}000$ to $100{,}000$, with $r=3$ and $\beta=6$. Second, we vary $r$ from $3$ to $50$, with $n=30{,}000$ and $\beta=3(r-1)$, to approximately maintain the expected proportion of within-community edges at $75\%$. Results are averaged over 10 independently generated networks for each configuration. Large sparse graphs are generated without storing the dense probability matrix. We adapt the geometric-skipping sampler of Miller and Hagberg~\cite{efficientgeneration} to each pair of communities: nodes are ordered by decreasing weight, and candidate edges are sampled using geometric jumps followed by an acceptance step. This procedure preserves independent Bernoulli sampling with probabilities $(Z\theta Z^\top)_{ij}$ for $i<j$. Accepted edges are stored directly in a sparse adjacency matrix. Figure~\ref{fig:scalability} reports the mean running time and RMI as functions of $n$ and $r$. When varying $n$, FROST's running time grows approximately linearly and remains below that of all DCBM inference methods considered, reaching approximately $30$ seconds at $n=100{,}000$, compared with $9$ minutes for graph-tool. Profiling shows that the abrupt increase in SCORE+'s running time originates in the $k$-means clustering step. When varying $r$, FROST's running time increases much more slowly than those of KN and KL-EM. It remains faster than graph-tool throughout the tested range, although the difference narrows as $r$ increases. Graph-tool's running time does not increase with $r$, consistent with the reported complexity of its multilevel agglomerative procedure, which is independent of the number of communities~\cite{peixoto2014efficient}. In this experiment, FROST and SCORE+ have comparable running times, with FROST consistently achieving higher RMI. SVCA initialization improves the recovery performance of KN throughout both experiments and also benefits KL-EM, particularly when varying $n$.

\paragraph{Robustness under edge addition}
\begin{figure}
    \centering
    \begin{tikzpicture}
\begin{axis}[
    width=0.95\linewidth, height=7.5cm,
    xlabel={$\varepsilon$}, ylabel={RMI},
    xmin=-0.03, xmax=1.03,
    xtick={0,0.1,0.2,0.4,0.6,0.8,1},
    grid=major, grid style={dotted,gray!50},
    legend style={
            at={(0.62,0.98)},
            anchor=north west,
            draw=none,
            font=\small,
            legend columns=2,
        },
    every axis plot/.append style={thick, mark size=2pt}
]

\addplot+[blue, solid, mark=*, mark options={fill=blue}] coordinates {
    (0,0.5101) (0.1,0.4423) (0.2,0.3907) (0.4,0.3146) (0.6,0.2469) (0.8,0.2095) (1,0.1750)
};
\addlegendentry{KN}

\addplot+[red, solid, mark=*, mark options={fill=red}] coordinates {
    (0,0.7013) (0.1,0.6201) (0.2,0.5626) (0.4,0.4636) (0.6,0.3949) (0.8,0.3309) (1,0.2796)
};
\addlegendentry{KL-EM}

\addplot+[blue, dashed, mark=square*, mark options={fill=blue,solid}] coordinates {
    (0,0.7018) (0.1,0.6271) (0.2,0.5704) (0.4,0.4842) (0.6,0.4207) (0.8,0.3732) (1,0.3335)
};
\addlegendentry{KN(S)}

\addplot+[red, dashed, mark=square*, mark options={fill=red,solid}] coordinates {
    (0,0.6996) (0.1,0.6303) (0.2,0.5719) (0.4,0.4859) (0.6,0.4221) (0.8,0.3707) (1,0.3319)
};
\addlegendentry{KL-EM(S)}

\addplot+[darkgreen, solid, mark=x] coordinates {
    (0,0.6945) (0.1,0.6165) (0.2,0.5472) (0.4,0.4451) (0.6,0.3381) (0.8,0.2520) (1,0.1581)
};
\addlegendentry{graph-tool}

\addplot+[cyan, solid, mark=star] coordinates {
    (0,0.6464) (0.1,0.6136) (0.2,0.5831) (0.4,0.5311) (0.6,0.4792) (0.8,0.4446) (1,0.4126)
};
\addlegendentry{SCORE+}

\addplot+[black, dashed, mark=triangle*, mark options={fill=black,solid}] coordinates {
    (0,0.6867) (0.1,0.6454) (0.2,0.6105) (0.4,0.5469) (0.6,0.4983) (0.8,0.4632) (1,0.4245)
};
\addlegendentry{FROST(S)}

\end{axis}
\end{tikzpicture}
    \caption{Average RMI over 100 sampled networks as a function of the perturbation level $\epsilon$.}
    \label{fig:edge perturbation}
\end{figure}

This experiment evaluates robustness to edge additions, simulating false-positive connections.  As in the previous experiments, we generate networks from the Bernoulli DCBM with $\beta=6$, $\gamma=2.5$, and $\alpha=10$.  Following~\cite{tian2023robustness}, we add edges uniformly at random among pairs of distinct, unconnected nodes. The perturbation level $\epsilon$ is the ratio of the number of added edges to the initial number of edges. Figure~\ref{fig:edge perturbation} shows the average RMI over 100 sampled networks for each perturbation level $\epsilon$ between $0$ and $1$. RMI decreases with $\epsilon$ for all methods. However, FROST and SCORE+ exhibit a smaller decline than the DCBM inference methods, including graph-tool, KN, and KL-EM, and the gap increases as more edges are added. SVCA initialization also improves the recovery performance of KN and KL-EM.
\subsection{Real-world networks}
We evaluate community recovery and sensitivity to initialization on four real networks commonly used as benchmarks for community detection.   These experiments complement synthetic tests by considering networks that may exhibit more complex structures and deviate from the assumptions of the DCBM model. The available labels serve as reference communities, even though they do not necessarily coincide exactly with the structural organization of the networks~\cite{peel2017ground}.

The \textbf{Karate} network~\cite{zachary} represents the social interactions among 34 members of a karate club at an American university. Following an internal conflict, the club split into two distinct factions.

The \textbf{Polblogs} network~\cite{Polblogs}  represents hyperlinks between blogs on U.S. politics, recorded in 2005. The reference labels distinguish liberal and conservative blogs. As in~\cite{SBM_degreecorection}, we ignore edge directions and retain the largest connected component, containing 1,222 nodes and 16,714 edges.

The \textbf{Football} network~\cite{football} represents games between U.S. college football teams during the 2000 regular season, with conference memberships serving as reference labels. Following~\cite{score+}, we remove the five independent teams, leaving 110 nodes, 570 edges, and 11 communities.

The \textbf{Caltech} network~\cite{Traud2012} represents Facebook friendships at the California Institute of Technology. The reference labels correspond to eight residential Houses. After removing nodes with missing residence information and retaining the largest connected component, the network contains 590 nodes and 12,822 edges.

\begin{table*}[t]
    \centering
    \small
    \setlength{\tabcolsep}{0pt}
    \renewcommand{\arraystretch}{1.15}
    \caption{RMI (\%) over 100 runs.
    Best obj.: RMI at the best objective value found across algorithms sharing the same objective,
    with each algorithm’s attainment frequency below.
    Mean: mean $\pm$ standard deviation of RMI.
    Bold indicates the highest RMI per network among the Best obj.\ results
    and the SCORE+ mean.
    (S) denotes SVCA initialization.}
    \label{tab:real_networks}
    \begin{tabular}{@{}lcccccccc@{}}
        \toprule
        & \multicolumn{2}{c}{Karate} & \multicolumn{2}{c}{Polblogs}
        & \multicolumn{2}{c}{Football} & \multicolumn{2}{c}{Caltech} \\
        \noalign{\vskip 2pt}
        \cline{2-9}
        \noalign{\vskip 2pt}
        Method & Best obj. & Mean & Best obj. & Mean
               & Best obj. & Mean & Best obj. & Mean \\
        \midrule
KN 
& \shortstack{$56.8$\\[-1pt]{\scriptsize (4\%)}} & $7.5\pm18.2$ 
&\shortstack{$72.1$\\[-1pt]{\scriptsize (7\%)}} & $22.6\pm32.4$ 
& \shortstack{$79.9$\\[-1pt]{\scriptsize (0\%)}} & $68.8\pm8.6$ 
& \shortstack{$62.1$\\[-1pt]{\scriptsize (0\%)}} & $52.5\pm6.3$ \\[3pt]
KN(S) 
& \shortstack{$56.8$\\[-1pt]{\scriptsize (33\%)}} & $44.0\pm23.7$ 
& \shortstack{$72.1$\\[-1pt]{\scriptsize (68\%)}} & $70.2\pm10.1$ 
& \shortstack{$79.9$\\[-1pt]{\scriptsize (0\%)}} & $81.8\pm4.4$ 
& \shortstack{$62.1$\\[-1pt]{\scriptsize (0\%)}} & $59.8\pm2.5$ \\[3pt]

KL-EM 
& \shortstack{$56.8$\\[-1pt]{\scriptsize (18\%)}} & $17.7\pm25.3$ 
& \shortstack{$72.1$\\[-1pt]{\scriptsize (15\%)}} & $56.9\pm28.3$ 
& \shortstack{$79.9$\\[-1pt]{\scriptsize (2\%)}} & $74.4\pm7.5$ 
& \shortstack{$62.1$\\[-1pt]{\scriptsize (1\%)}} & $56.0\pm5.3$ \\[3pt]
KL-EM(S) 
& \shortstack{$56.8$\\[-1pt]{\scriptsize (39\%)}} & $45.4\pm22.8$ 
& \shortstack{$72.1$\\[-1pt]{\scriptsize (68\%)}} & $70.9\pm7.2$ 
& \shortstack{$79.9$\\[-1pt]{\scriptsize (0\%)}} & $81.8\pm4.5$ 
& \shortstack{$62.1$\\[-1pt]{\scriptsize (0\%)}} & $59.8\pm2.5$ \\[3pt]

FROST(S) 
& \shortstack{$\mathbf{74.5}$\\[-1pt]{\scriptsize (74\%)}} & $55.3\pm32.6$ 
& \shortstack{$70.1$\\[-1pt]{\scriptsize (70\%)}} & $68.0\pm12.0$ 
& \shortstack{$86.5$\\[-1pt]{\scriptsize (25\%)}} & $80.2\pm5.3$ 
& \shortstack{$\mathbf{65.7}$\\[-1pt]{\scriptsize (1\%)}} & $59.3\pm3.6$ \\[3pt]

graph-tool 
& \shortstack{$0.0$\\[-1pt]{\scriptsize (17\%)}} & $39.2\pm35.2$ 
& \shortstack{$71.6$\\[-1pt]{\scriptsize (5\%)}} & $70.8\pm1.0$ 
& \shortstack{$\mathbf{92.4}$\\[-1pt]{\scriptsize (24\%)}} & $88.6\pm3.5$ 
& \shortstack{$65.5$\\[-1pt]{\scriptsize (1\%)}} & $57.7\pm2.8$ \\[3pt]

SCORE+ & N/A & $62.3\pm0.0$
       & N/A & $\mathbf{74.4}\pm0.0$
       & N/A & $83.7\pm0.0$
       & N/A & $63.2\pm0.4$  \\[3pt]
        \bottomrule
    \end{tabular}
\end{table*}

We perform 100 runs per method for each network. Table~\ref{tab:real_networks} reports the RMI of the best solution found across all runs and algorithms optimizing that objective, together with each algorithm’s frequency of attaining this value. It also reports the mean and standard deviation
of RMI over individual runs. FROST obtains partitions that agree well with the reference labels. SVCA initialization improves the mean RMI of both KN and KL-EM on all four networks and often increases the frequency of reaching their best observed objective. These results highlight the benefits of SVCA initialization and the value of multiple restarts, retaining the solution with the best objective value.

\begin{figure}[b!]
    \centering

    \begin{subfigure}{\linewidth}
        \centering
        \includegraphics[width=8cm]{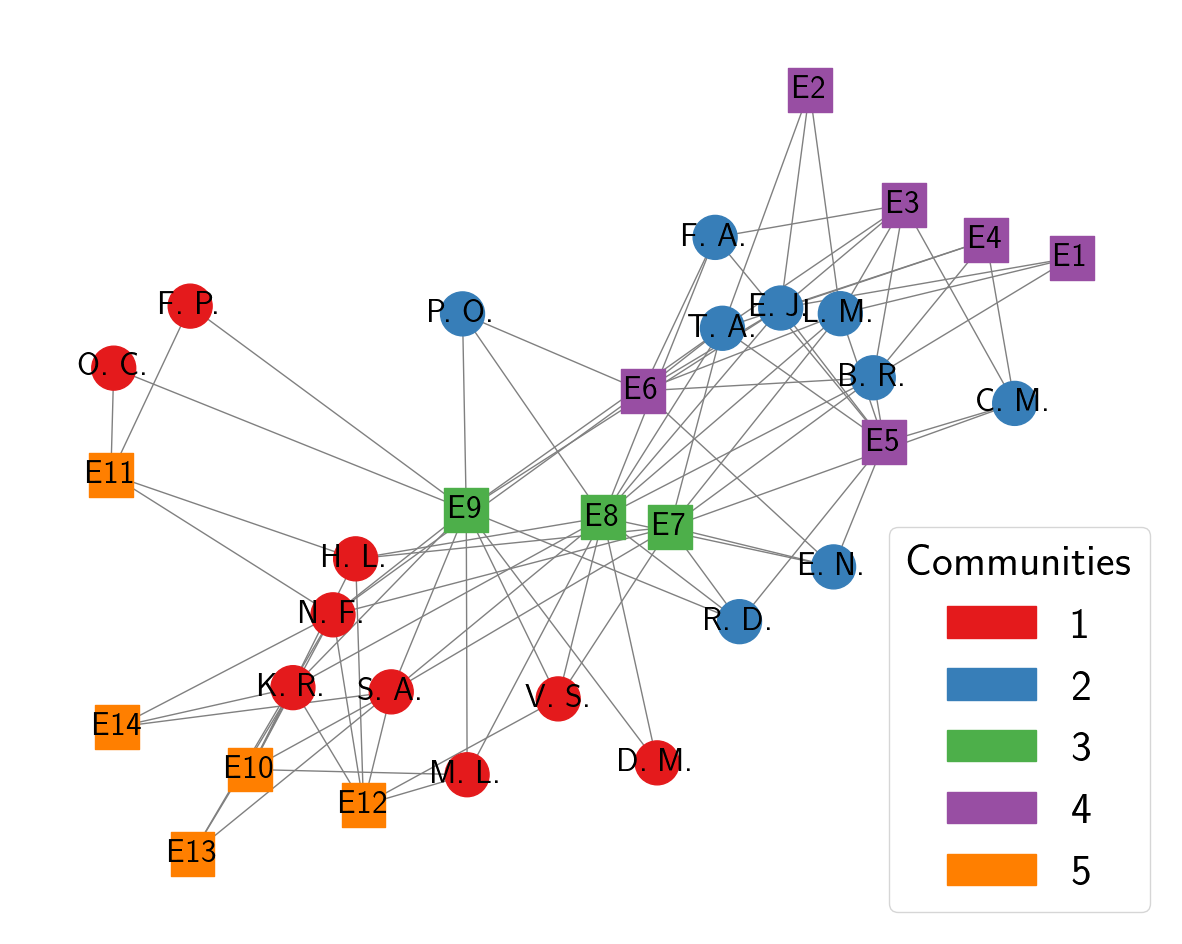}
        \caption{The five communities identified using OtrisymNMF in the Southern women network, with women (circles) and events (squares) clearly separated, and each community represented by a distinct color.}
        \label{fig:southernwomen_otr}
    \end{subfigure}

    \vspace{0.5em}

    \begin{subfigure}{\linewidth}
        \centering
        \includegraphics[width=3.6cm]{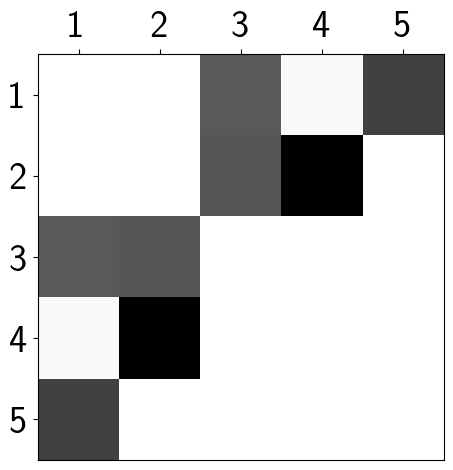}
        \caption{Heatmap of the matrix $\theta$ found using OtrisymNMF, illustrating interactions between communities.}
        \label{fig:southernS}
    \end{subfigure}

    \caption{Results obtained with OtrisymNMF on the Southern women network.}
    \label{fig:southern_results}
\end{figure}

To illustrate the ability of the models to recover bipartite structures, we also consider the Southern women network~\cite{southernwomen}, which records the attendance of women at social events held in a southern town in the United States. The bipartite network is composed of 32 nodes, 18 for women and 14 for events. An edge exists between a woman and an event if the woman attended that event. Following~\cite{SBM&DCBMforBIPARTITE}, we use $r=5$. OtrisymNMF and DCBM identify the same partition shown in Fig.~\ref{fig:southernwomen_otr}. This partition perfectly matches the one found by the biSBM~\cite{SBM&DCBMforBIPARTITE}, with the partition of women aligning perfectly with the consensus in the literature~\cite{SBM&DCBMforBIPARTITE}. Fig.~\ref{fig:southernS} shows the matrix $\theta$, clearly illustrating the bipartition and the interactions between communities.

 \section{Conclusion}
In this paper, we related the degree-corrected block model (DCBM) to nonnegative matrix factorization (NMF). Specifically, maximum-likelihood estimation under the DCBM of Karrer and Newman~\cite{SBM_degreecorection} is equivalent to minimizing the Kullback-Leibler (KL) divergence between the adjacency matrix and a constrained nonnegative trifactorization. We investigated OtrisymNMF as an alternative that retains the same factorization and constraints but uses a Frobenius reconstruction error. To account for degree-dependent variation in noise, we applied this formulation to the regularized normalized adjacency matrix, corresponding to a weighted Frobenius error on the original adjacency matrix.

We presented FROST, a block-coordinate descent algorithm for OtrisymNMF, together with a theoretically grounded initialization based on smoothed separable NMF. This initialization exploits the separability of the model and uses SVCA~\cite{nadisic2023smoothed} to estimate the factors while reducing sensitivity to noise. It provides initial factors for FROST and initial node partitions for DCBM inference algorithms, without assuming a particular type of block structure. 

Experiments on synthetic and real networks compared FROST with DCBM inference methods (KN, KL-EM and graph-tool), with SCORE+ as an additional spectral baseline. On synthetic networks, FROST achieved the best community recovery among the tested methods in several challenging settings, including weakly separated communities, disassortative structures, sparse graphs, and strong degree heterogeneity. It also showed greater robustness to random edge additions than the DCBM inference methods considered. The real-network results further support OtrisymNMF with FROST as a competitive alternative to DCBM inference. In the scalability experiments, FROST was faster than the DCBM inference methods considered. It processed networks with 100,000 nodes and approximately 500,000 edges in about 30 seconds, compared with 9 minutes for graph-tool. SVCA initialization also improved the recovery performance of KN and KL-EM compared with random initialization, demonstrating its usefulness beyond FROST.

The connection between DCBM inference and NMF also clarifies the interpretation of OtrisymNMF in terms of community memberships, degree heterogeneity, and interactions between communities. It motivates exploring other reconstruction errors within the same constrained trifactorization framework and applying further theoretical and algorithmic developments from NMF to DCBM inference and community detection.

\section{Acknowledgments}
This research is funded by the European Union (ERC consolidator, eLinoR, no 101085607). Alexandra Dache is a Research Fellow of the Fonds de la Recherche Scientifique - FNRS (F.R.S.-FNRS).

\bibliographystyle{IEEEtran}
\bibliography{biblio}
\end{document}